\documentclass[reprint,amssymb,amsmath,superscriptaddress,aps,pra,nofootinbib]{revtex4-1}

\usepackage{color}
\usepackage{bm}
\usepackage{amsmath,amssymb}
\usepackage{amsfonts}
\usepackage{dsfont}
\usepackage{graphicx}
\usepackage{float}
\usepackage{subfigure} 
\usepackage{verbatim}
\usepackage{amsfonts}
\usepackage{bbold}
\usepackage{dcolumn}
\usepackage{bm}
\usepackage{color}
\usepackage[dvipsnames]{xcolor}
\usepackage{listings}
\definecolor{blueprl}{RGB}{46,48,146}
\usepackage[colorlinks=true,urlcolor=blueprl,citecolor=blueprl,linkcolor=blueprl]{hyperref}
\usepackage{mathrsfs}
\usepackage{filecontents}
\usepackage{mathtools}
\usepackage{times} 

\newcommand{\ket}[1]{\mbox{$| #1 \rangle$}}

\newcommand{\tr}{\mbox{tr}}

\newcommand{\xdownarrow}[1]{%
  {\left\downarrow\vbox to #1{}\right.\kern-\nulldelimiterspace}
}

\begin{document}

\title{Teleportation-based collective attacks in Gaussian quantum key distribution}
\author{Spyros Tserkis}  \email{spyrostserkis@gmail.com}
\affiliation{Centre for Quantum Computation and Communication Technology, School of Mathematics and Physics, University of Queensland, St Lucia, Queensland 4072, Australia}
\affiliation{Centre for Quantum Computation and Communication Technology, Department of Quantum Science, Australian National University, Canberra, ACT 2601, Australia.}
\author{Nedasadat Hosseinidehaj} 
\affiliation{Centre for Quantum Computation and Communication Technology, School of Mathematics and Physics, University of Queensland, St Lucia, Queensland 4072, Australia}
\author{Nathan Walk} 
\affiliation{Dahlem Center for Complex Quantum Systems, Freie Universit\"{a}t Berlin, 14195 Berlin, Germany}
\author{Timothy C. Ralph}
\affiliation{Centre for Quantum Computation and Communication Technology, School of Mathematics and Physics, University of Queensland, St Lucia, Queensland 4072, Australia}
\date{\today}

\begin{abstract}
In Gaussian quantum key distribution eavesdropping attacks are conventionally modeled through the universal entangling cloner scheme, which is based on the premise that the whole environment is under control of the adversary, i.e., the eavesdropper purifies the system. This assumption implies that the eavesdropper has either access to an identity (noiseless) channel or infinite amount of entanglement in order to simulate such an identity channel. In this work we challenge the necessity of this assumption, and we propose a teleportation-based eavesdropping attack, where the eavesdropper is not assumed to have access to the shared channel, that represents the unavoidable noise due to the environment. Under collective measurements, this attack reaches optimality in the limit of infinite amount of entanglement, while for finite entanglement resources it outperforms the corresponding optimal individual attack. We also calculate the minimum amount of distributed entanglement that is necessary for this eavesdropping scheme, since we consider it as the operationally critical quantity capturing the limitations of a realistic attack. We conclude that the fact that infinite amount of entanglement is required for an optimal collective eavesdropping attack signifies the robustness of Gaussian quantum key distribution.
\end{abstract}

\maketitle

\section{Introduction}

Quantum Key Distribution (QKD) \cite{Scarani.et.al.RMP.09,Pirandola.et.al.arxiv.19,Xu.et.al.arxiv.19} is one of the most prominent quantum communication protocols, which enables two parties (Alice and Bob) to establish a shared (random) secret key for cryptographic purposes. It was originally developed for discrete-variable (DV) quantum systems \cite{Bennett.Brassard.IEEE.84,Ekert.PRL.91}, but has also been extended to the continuous-variable (CV) regime \cite{Ralph.PRA.99,Hillery.PRA.00,Reid.PRA.00}. A clear advantage of the CV-QKD schemes (see Refs.~\cite{Jouguet.el.al.PRA.11,Jouguet.et.al.NP13,Huang.et.al.SR.16,Zhang.et.al.QST.19,Zhang.et.al.NP.19} for recent advances) over their DV counterparts is the low-cost telecom optical components needed, which are already available for classical communication.

We generally consider the following assumptions for an eavesdropper (Eve) in a QKD scheme \cite{Scarani.et.al.RMP.09,Pirandola.et.al.arxiv.19,Xu.et.al.arxiv.19}: (i) Eve has full access to the quantum channel between Alice and Bob, (ii) Eve has unlimited computational power, (iii) Eve can monitor the public classical channel, but she cannot modify the messages (authenticated channel), and (iv) Eve has no access to Alice's and Bob's laboratories.

The most powerful attack Eve can asymptotically perform is the so-called collective attack \cite{Renner.PhD.05}, where she prepares and interacts a set of individual and identical quantum systems with the quantum signals sent from Alice to Bob. She then stores the output ensemble into her quantum memory for a future collective measurement (an individual attack would rely on individual measurements respectively). Unconditional security of a QKD protocol can be achieved by upper bounding the information Eve can extract, also known as the Holevo bound \cite{Holevo.PIT.73}.

The question we answer in the context of Gaussian QKD in this paper is the following: \textit{can Eve optimally attack the system under collective measurements without having access to the quantum channel between Alice and Bob?} In other words, we investigate the necessity of the first assumption (discussed before) regarding Eve's capabilities. We conclude that \textit{it is indeed not necessary for Eve to have access to the channel in order to collectively attack a system as long as she can perform an all-optical teleportation \cite{Ralph.OL.99} over it}. In particular, we propose a teleportation-based eavesdropping scheme that serves as an alternative type of attack to the well-known entangling cloner \cite{Grosshans.et.al.Nature.03,Grosshans.et.al.QIC.03,Pirandola.Braunstein.Lloyd.PRL.08} that assumes Eve's access to the shared quantum channel.

We further discuss how, under this scheme, Eve's information depends on the amount of entanglement she can prepare, distribute and distill in order to successfully perform the all-optical teleportation protocol \cite{Ralph.OL.99} (see Refs.\cite{Bennett.et.al.PRL.93,Brassard.Braunstein.Cleve.PD.98,Vaidman.PRA.94,Braunstein.Kimble.PRL.98,Braunstein.et.al.PRL.00,Andersen.Ralph.PRL.13,Marshall.James.JOSAB.14,Pirandola.et.al.NP.15}). Employing collective measurements, and using a resource state with the least required amount of entanglement, Eve's information reaches the bound for optimal individual attacks \cite{Lodewyck.Grangier.PRA.07,Sudjana.et.al.PRA.07}. Using the same setup and taking the limit to infinite entanglement for the resource state, Eve's information approaches the ultimate bound for an eavesdropping attack, also known as the Holevo bound \cite{Holevo.PIT.73}.

We identify the distributed entanglement used as a resource for teleportation as the operationally critical quantity capturing the limitations of a realistic Eve. Under this limitation we evaluate the secret key rate that Alice and Bob achieve conditioned on their belief on how powerful (in entanglement resources) Eve is. Note that other types of physical limitations on Eve's capabilities for CV-QKD systems have also been studied in Refs.~\cite{Hosseinidehaj.Walk.Ralph.PRA.19,Pan.et.al.arXiv.19}.

In Sec.~\ref{secgcvqkd} we briefly introduce how CV-QKD works in the Gaussian regime, and in Sec.~\ref{secueca} we present the conventional way an eavesdropping attack is modeled through the universal entangling cloner scheme. The alternative eavesdropping attack based on the all-optical teleportation is proposed in Sec.~\ref{secaota}, and in Sec.~\ref{secd} we discuss the results of a specific example. Finally, we conclude this work with Sec.~\ref{secc}. The protocols of both the standard CV teleportation and the all-optical teleportation are discussed in the App.~\ref{seca}.

\section{Gaussian CV-QKD}
\label{secgcvqkd}

A generic QKD protocol in prepare-and-measure (PM) scheme consists of: (i) quantum communication, where Alice encodes classical information into conjugate quantum basis states, which are sent through an insecure quantum channel to Bob, who measures the received quantum states in a randomly chosen basis, resulting in two sets of correlated data, and (ii) classical communication (classical post-processing) over a public but authenticated classical channel, where Alice and Bob extract a secret key from the correlated data they collected during the previous step.

In a fully Gaussian CV-QKD protocol \cite{Garcia-Patron.PhD.07,Weedbrook.et.al.RVP.12} (in the PM scheme) Alice encodes a classical random variable ``a" (drawn from a Gaussian distribution) onto Gaussian quantum states, squeezed states \cite{Cerf.Levy.VanAssche.PRA.01} or coherent states \cite{Grosshans.Grangier.PRL.02}, and sends them through an insecure quantum channel to Bob, who measures the received quantum states using homodyne or heterodyne detection to obtain a classical random variable ``b".  

Gaussian (collective or individual) attacks are asymptotically optimal \cite{Renner.Cirac.PRL.2009,Garcia-Patron.Cerf.PRL.06,Navascues.Grosshans.PRL.06,Leverrier.Grangier.PRA.10}, and the asymptotic secret key rate against optimal collective attacks is given by \cite{Renner.Gisin.Kraus.PRA.05,Devetak.Winter.PRSA.05}
\begin{equation}
K := \beta I(\text{a}{:}\text{b})- \mathcal{S}(\text{x}{:}E) \,,
\label{keyrate}
\end{equation}
where $I(\text{a}{:}\text{b})$ is the classical mutual information between Alice and Bob, and $\mathcal{S}(\text{x}{:}E)$ is the maximum mutual information between Alice ($\text{x}{\equiv}\text{a}$) and Eve (in the direct reconciliation where Alice is the reference of the reconciliation in the classical post-processing), or between Bob ($\text{x}{\equiv}\text{b}$) and Eve (in the reverse reconciliation where Bob is the reference of the reconciliation). The coefficient $0 \leqslant \beta \leqslant 1$ is the reconciliation efficiency \cite{Jouguet.el.al.PRA.11,Jouguet.et.al.NP13,Huang.et.al.SR.16,Zhang.et.al.QST.19,Zhang.et.al.NP.19}. Note that the maximum amount of information Eve can possibly extract from the collective attack is upper bounded by the Holevo bound $\chi(\text{x}{:}E)$ \cite{Holevo.PIT.73}, i.e., 
\begin{equation}
\mathcal{S}(\text{x}{:}E) \leqslant \chi(\text{x}{:}E) \,.
\label{holevobound}
\end{equation}

Gaussian states \cite{Holevo.B.19,Adesso.Ragy.OSID.14,Weedbrook.et.al.RVP.12,Serafini.B.17} $\hat{\sigma}$ are the ones that can be fully characterized by the mean value and the variance of the quadrature field operators $\hat{q}:=(\hat{x}_{1},\hat{p}_{1},\ldots
,\hat{x}_{n},\hat{p}_{n})^{T}$, with $\hat{x}_{j}:=\hat{a}_{j}+\hat{a}%
_{j}^{\dag}$ and $\hat{p}_{j}:=i(\hat{a}_{j}^{\dag}-\hat{a}_{j})$, where
$\hat{a}_{j}$ and $\hat{a}_{j}^{\dag}$ are the annihilation and creation
operators, respectively. Without losing generality we assume a zero mean-valued state that can be fully described by its covariance matrix, whose arbitrary element is given by $\sigma_{ij}:=\frac{1}{2}\langle \{\hat{q}_{i},\hat{q}_{j}\}\rangle$. The covariance matrix in the standard form \cite{Duan.et.al.PRL.00,Simon.PRL.00} is given by 
\begin{equation}
\boldsymbol{\sigma}^{\text{sf}}=\begin{bmatrix}
\boldsymbol{A} & \boldsymbol{C} \\
\boldsymbol{C} & \boldsymbol{B}
\end{bmatrix}\,,
\label{cov}
\end{equation}
with $\boldsymbol{A} = \text{diag}(a,a)$, $\boldsymbol{B} = \text{diag}(b,b)$ and $\boldsymbol{C} = \text{diag}(c_+,c_-)$, where $a \geqslant b$ and $c_+\geqslant |c_-| \geqslant 0$. A two-mode squeezed vacuum has $a=b=\frac{1+\zeta^2}{1-\zeta^2}$ and $c_+=-c_-=\frac{2 \zeta }{1-\zeta ^2}$, where $0 \leqslant \zeta < 1$ is the squeezing parameter. The covariance matrix transformation when a phase-insensitive single-mode Gaussian channel $\mathcal{G}$ acts on one arm of a two-mode Gaussian state $\boldsymbol{\sigma}_{\text{in}}$ is given by \cite{Holevo.PIT.07}
\begin{equation}
\boldsymbol{\sigma}_{\text{out}} = \mathcal{G}(\boldsymbol{\sigma}_{\text{in}})=(\mathds{1} \oplus U)\boldsymbol{\sigma}_{\text{in}} (\mathds{1} \oplus U)^T + (\mathbb{0} \oplus V) \,,
\label{channel}
\end{equation}
where $U= \sqrt{\tau} \mathds{1}$ and $V= v \mathds{1}$. Significant phase-insensitive Gaussian channels are the following: (i) the lossy channel $\mathcal{L}$ with transmissivity $0 < \tau < 1$ and noise $v = (1-\tau)\epsilon$ (pure loss $\mathcal{L}_p$ for $\epsilon=1$, thermal loss for $\epsilon >1$), (ii) the amplifier channel $\mathcal{A}$ with gain $\tau > 1$ and noise $v = (\tau-1)\epsilon$ (pure amplifier $\mathcal{A}_p$ for $\epsilon=1$, thermal amplifier for $\epsilon >1$), (iii) the classical additive noise channel $\mathcal{N}$ with $\tau= 1$ and noise $v>0$, and (iv) the identity channelI with $\tau= 1$ and $v=0$, representing the ideal nondecohering channel.

Let us assume that Alice and Bob identify in their in-between interaction a phase-insensitive channel $\mathcal{G}$, that we assume to be a thermal lossy channel with transmissivity $0 < \tau < 1$ and noise $v=(1-\tau)\varepsilon$, but the result can be trivially extended to any non-entanglement-breaking \cite{Namiki.Hirano.PRL.04} phase-insensitive channel.

Each Gaussian PM scheme can be represented using an equivalent entanglement-based scheme \cite{Garcia-Patron.PhD.07}, where Alice prepares a pure Gaussian entangled state, i.e., a two-mode squeezed vacuum state $\boldsymbol{\sigma}_{\text{in}}$, keeping one mode while sending the second mode through the quantum channel. If Alice applies a homodyne (heterodyne) detection, the second mode of the entangled states is projected onto a squeezed (coherent) state. While in the experimental demonstration of CV-QKD, PM scheme is preferred, the entanglement-based scheme is favored for the security analysis.  

\section{Universal Entangling Cloner Attack}
\label{secueca}

\begin{figure}[t]
\centering
  \includegraphics[width=\columnwidth]{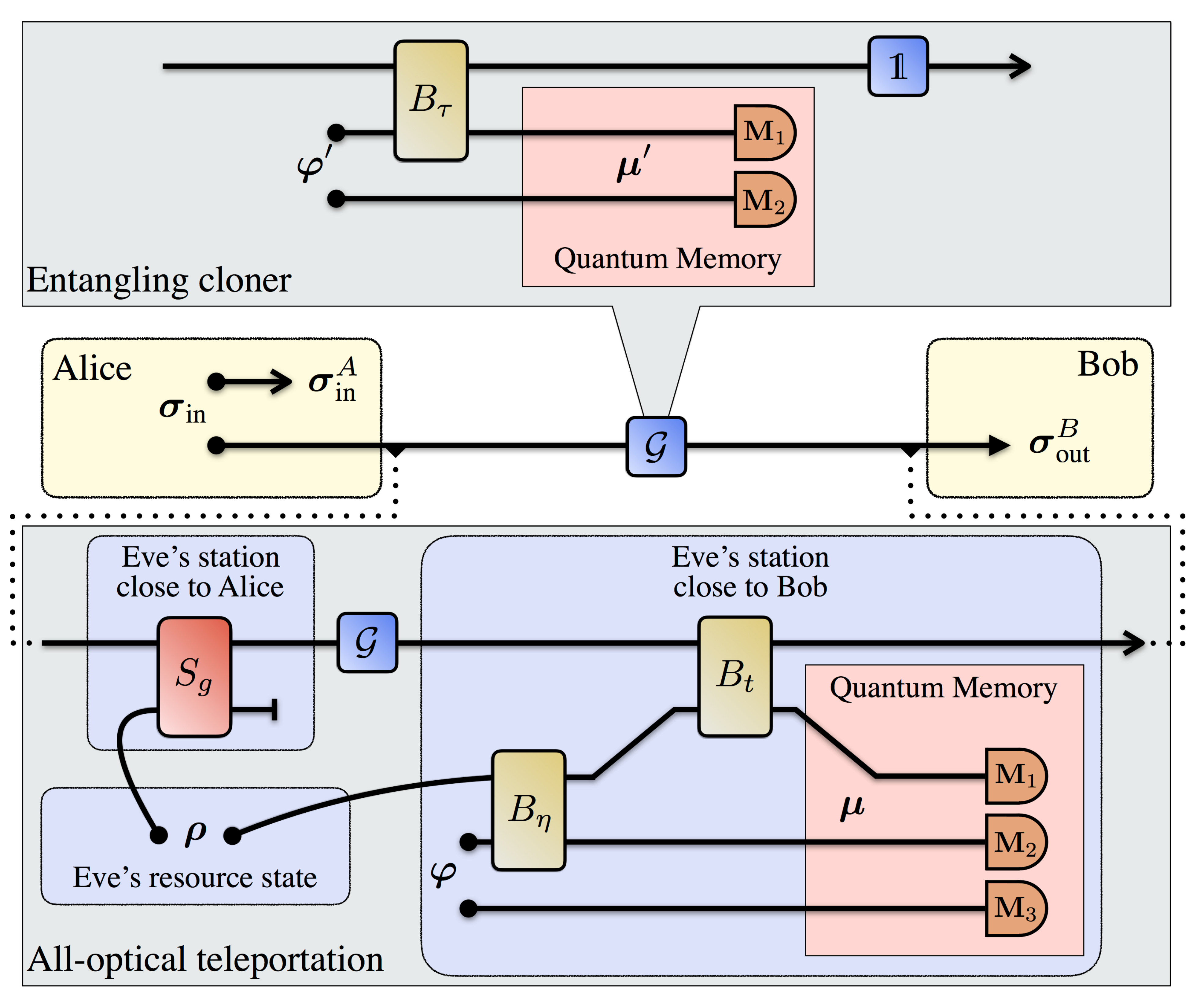}
  \caption{Eavesdropping attack. On the top panel we present the entangling cloner attack, where Eve simulates the channel $\mathcal{G}$ through a beam-splitter $B_{\tau}$ (with the same transmissivity as the channel $\mathcal{G}$). One input of the beam-splitter is Alice's signal and the other is one arm of Eve's state $\boldsymbol{\varphi}'$. With $\text{M}_{1-2}$ we represent the measurements that Eve performs later on her quantum memory on her state $\boldsymbol{\mu}'$, and $\mathds{1}$ the identity channel. On the bottom panel, we present the all-optical teleportation attack, where Eve performs an all-optical teleportation over the signal sent from Alice. With $\boldsymbol{\rho}$ we denote Eve's distilled resource state. The all-optical teleportation consists of a two-mode squeezer $S_g$ with gain $g$, which is in Eve's first station close to Alice, and a beam-splitter $B_t$ with transmissivity $t=1/g$, which is in Eve's second station close to Bob. One mode of Eve's resource state $\boldsymbol{\rho}$ is sent to the first station as an input of $S_g$. The other mode of $\boldsymbol{\rho}$ is sent to the second station and is mixed on a beam-splitter $B_{\eta}$ with another state $\boldsymbol{\varphi}$, before it becomes an input to $B_t$. Finally Eve performs the measurements $\text{M}_{1-3}$ on her state $\boldsymbol{\mu}$ that was stored in the quantum memory.}
  \label{fig1}
\end{figure}

In the entangling cloner setup \cite{Grosshans.et.al.Nature.03,Grosshans.et.al.QIC.03,Pirandola.Braunstein.Lloyd.PRL.08}, the whole channel $\mathcal{G}$ is associated with a potential eavesdropper that has full control of the environment. Eve uses a two-mode squeezed vacuum state and mixes one arm of it with Alice's signal in a beam-splitter with transmissivity equal to the transmissivity of the quantum channel. For the collective attack, one of the outputs is sent directly to Bob while the rest are stored in her quantum memory. Finally, she collectively measures the stored ensemble to gain the maximum information about the distributed key. This scheme is schematically represented in Fig.~\ref{fig1}.

An assumption that has been taken in this setup is that Eve is able to noiselessly transmit the output signal of the beam-splitter to Bob. Obviously, this is a really strong assumption, since Eve has to deal with some unavoidable decoherence due to environmental reasons that go beyond her control, but even theoretically that is impossible because simulating identity channels through teleportation in CV systems requires an infinite amount of entanglement.

For the case of optimal individual attacks it has already been shown \cite{Lodewyck.Grangier.PRA.07, Sudjana.et.al.PRA.07} that they can be realistically modeled through the standard CV-teleportation protocol \cite{{Braunstein.Kimble.PRL.98}}, however this protocol is dependent on individual Bell-type measurements, and thus it cannot be directly used for collective attacks. In order to realistically model an optimal collective attack, we propose below an eavesdropping scheme, based on the all-optical teleportation protocol \cite{Ralph.OL.99} that is measurement-free. 

\section{All-optical Teleportation Attack}
\label{secaota}

In this type of attack, represented in Fig.~\ref{fig1}, we start by assuming that there exists a physical quantum channel, $\mathcal{G}$, between Alice and Bob, through which all the participants (including Eve) must send their signals. Given this limitation, Eve (who is allowed to establish stations arbitrarily close to Alice's and Bob's laboratories) performs an all-optical teleportation protocol \cite{Ralph.OL.99} (details can be found in the App.~\ref{seca}) over this channel $\mathcal{G}$. In general, any Gaussian channel can be simulated via a quantum teleportation protocol using an appropriate resource state \cite{Giedke.Cirac.PRA.02,Niset.Fiurasek.Cerf.PRL.09,Pirandola.et.al.NC.17}. Given a Gaussian phase-insensitive channel $\mathcal{G}$ the set of all resource states that can simulate it have been derived in Ref.~\cite{Tserkis.Dias.Ralph.PRA.18} (see also Ref.~\cite{Scorpo.et.al.PRL.17}). 

For our purposes, we assume that Eve can prepare, distribute and distill a pure two-mode squeezed vacuum state $\boldsymbol{\rho}$ with squeezing parameter $0 \leqslant \gamma < 1 $. One arm of it is used for the initial amplification (performed on Eve's station close to Alice's side) through the two-mode squeezer $S_g$ with gain $g>1$ \cite{Caves.PRD.82}. Employing a beam-splitter $B_{\eta}$ with transmissivity $0 \leqslant \eta \leqslant 1$, we mix the second arm of $\boldsymbol{\rho}$ with one arm of another two-mode squeezed vacuum state $\boldsymbol{\varphi}$ with squeezing parameter $0 \leqslant \kappa < 1$, that is prepared and used on Eve's station close to Bob's side (note that for a pure loss channel $\mathcal{G}$ the two-mode squeezed vacuum states $\boldsymbol{\varphi}$ and $\boldsymbol{\varphi}'$ are reduced to single-mode vacuum states $\ket{0}$ for both the entangling cloner and the all-optical teleportation attack). One output of $B_{\eta}$ is headed to another beam-splitter $B_t$ with transmissivity $t=1/g$ for the final attenuation of the signal before it is forwarded to Bob.

The final step for Eve is to perform a collective measurement on the modes that she has stored in her quantum memory, denoted as a quantum state $\boldsymbol{\mu}$. The maximum information that she can extract from those measurements is given by \cite{Garcia-Patron.PhD.07}
\begin{equation} 
\mathcal{S}(\text{x}{:}E) = \mathcal{S}(\boldsymbol{\mu}) - \mathcal{S}(\boldsymbol{\mu}|\text{x}) \,,
\label{Eve-info}
\end{equation}
where $\mathcal{S}(\cdot)$ denotes the von Neumann entropy, that can be calculated through the symplectic eigenvalues $\nu_i$ of a $N$-mode state \cite{Weedbrook.et.al.RVP.12,Adesso.Ragy.OSID.14,Holevo.Sohma.Hirota.PRA.99}, via
\begin{equation} 
\mathcal{S}(\boldsymbol{\sigma}):= \sum_{i=1}^N \frac{\nu_i+1}{2} \log_2 \frac{\nu_i+1}{2} - \frac{\nu_i-1}{2} \log_2 \frac{\nu_i-1}{2} \,,
\end{equation}
and $\mathcal{S}(\boldsymbol{\mu}|\text{x})$ is the von Neumann entropy for Eve's quantum system conditioned on Alice or Bob's measurement.

The challenging part for Eve is to prepare, distribute and distill pure entangled states $\boldsymbol{\rho}$ over the corresponding distance between Alice and Bob. Thus, we assess her performance through the amount of entanglement of her state $\boldsymbol{\rho}$. The least amount of entanglement needed for the simulation of the channel $\mathcal{G}$ \cite{Tserkis.Dias.Ralph.PRA.18}, and consecutively for any teleportation-based attack is given by   
\begin{equation}
\mathcal{E}(\boldsymbol{\rho}) \geqslant \mathcal{E}(\gamma_{\text{min}}) \,,
\label{bound}
\end{equation}
where
\begin{equation}
\gamma_{\text{min}}=\frac{2 \sqrt{\tau }-\sqrt{(v+1-\tau) ( v-1+\tau)}}{\tau +v+1} \,,
\label{minsq}
\end{equation}
with $\mathcal{E}$ being the entropy of entanglement \cite{Bennett.DiVincenzo.et.al.PRA.96}, given by
\begin{equation}
\mathcal{E}(\gamma):=\frac{2 \gamma ^2 \log_2 \gamma+\left(1-\gamma ^2\right) \log_2 (1-\gamma ^2)}{\left(\gamma ^2-1\right) \ln2} \,.
\label{entropyofentanglement}
\end{equation}

\begin{figure*}[!htb]
  \centering
  \subfigure{\includegraphics[scale=0.25]{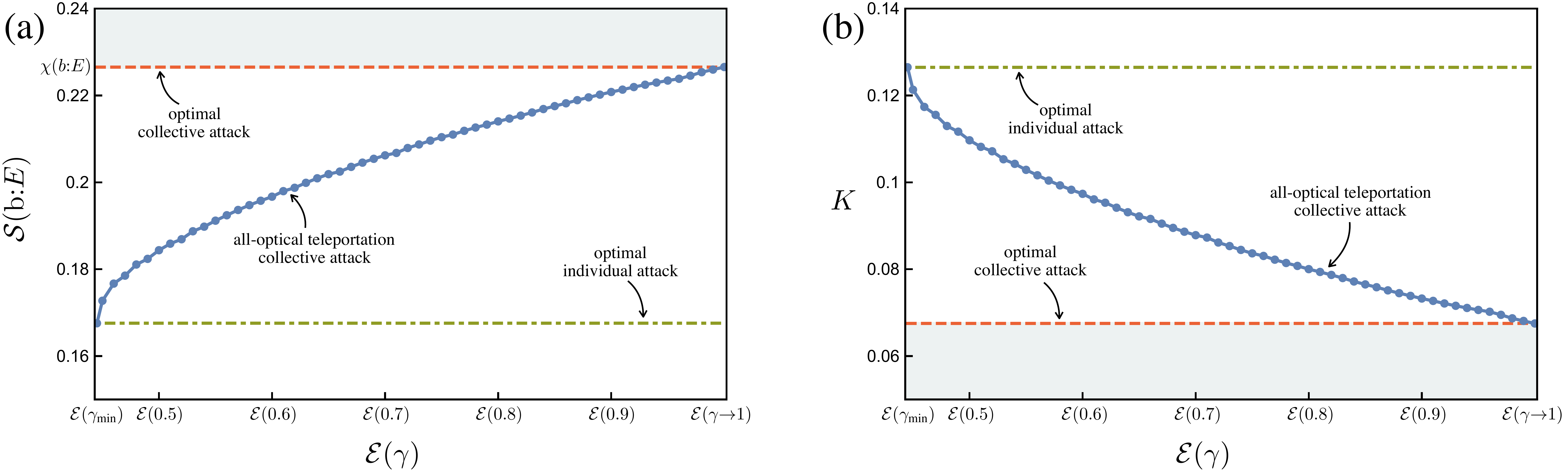}}
  \caption{ \small Eve's information and key rate. In figure (a) with the solid blue line we plot the amount of information $\mathcal{S}(\text{b}{:}E)$ Eve can extract in the reverse reconciliation scenario against the entanglement of her pure resource state $\mathcal{E}(\boldsymbol{\rho})$, parametrized over the squeezing parameter $\gamma$. The horizontal red dashed line represents the Holevo bound $\chi(b{:}E)$, and in the limit of infinite entanglement, i.e., $\mathcal{E}(\gamma \rightarrow 1) \rightarrow \infty$, we see that this bound is reached. With the green dot-dashed line we indicate the maximum amount of information Eve can extract through an optimal individual attack \cite{Lodewyck.Grangier.PRA.07, Sudjana.et.al.PRA.07}. In figure (b) we have the corresponding key rate that Alice and Bob measure given Eve's collective attack. Again, the red dashed line is the minimum possible key rate that they can extract, and the green dot-dashed represents the key rate based on the optimal individual attack. In both plots grey color indicates non-physical areas.}
\label{fig2}
\end{figure*}

\section{Discussion} 
\label{secd}

For our numerical calculations we assume that the quantum channel that Alice and Bob initially share is a thermal loss channel $\mathcal{G}$ with $\tau=0.25$, corresponding to approximately 30km of optical fiber, and $\varepsilon=1.01$. Alice's state is a two-mode squeezed vacuum with squeezing parameter $\zeta=0.7$, and the reconciliation efficiency is set equal to $\beta=0.95$. Alice and Bob measure their modes with heterodyne detection, and Eve performs an all-optical teleportation attack in the limit of $g \rightarrow \infty$ (the protocol works for finite amount of gain $g$ as well but with less success for Eve).

The plot in Fig.~\ref{fig2} (a) shows Eve's information for this protocol in the reverse reconciliation scenario, maximized over the parameters $\{\eta,\kappa \}$ that can simulate the channel $\mathcal{G}$, as a function of the entanglement of state $\boldsymbol{\rho}$. Note that for a pure loss channel there is no need for an optimization, since the transmissivity can be just set equal to $\eta=\tau/\gamma^2$. The value for the Holevo bound that Eve has to reach is given by \cite{Garcia-Patron.PhD.07} 
\begin{equation}
\chi= \mathcal{S}(\boldsymbol{\sigma}_{\text{out}}) - \mathcal{S}(\boldsymbol{\sigma}_{\text{out}}|\text{b}) \,,
\end{equation}
where $\mathcal{S}(\boldsymbol{\sigma}_{\text{out}}|\text{b})$ is the von Neumann entropy of the entangled state shared between Alice and Bob conditioned on Bob's measurement. The plot in Fig.~\ref{fig2} (b) shows the achievable secret key rate calculated by Alice and Bob, based on Eq.~(\ref{keyrate}) and the mutual information, which for this protocol is given by
\begin{equation}
I(\text{a}{:}\text{b})=\log_2  \frac{a \tau +v+1}{\tau +v+1} \,.
\label{mutual}
\end{equation}

The key rate calculated in Fig.~\ref{fig2} is for the asymptotic regime. However, if we include the finite-size effects \cite{Leverrier.PRL.15,Leverrier.PRL.17}, there would be some circumstances that while positive finite key rates cannot be generated from optimal collective attacks [under the unrealistic assumption of infinite entanglement with $\mathcal{E}(\gamma {\rightarrow} 1)$], by considering optimal individual attacks [under the assumption that Eve can distill a pure entangled state with $\mathcal{E}(\gamma_{\text{min}})$] we are able to move from insecure regime to secure regime, and generate non-trivial positive finite key rates.

We also plot the corresponding values for Eve's information and secret key rate under an optimal individual attack \cite{Lodewyck.Grangier.PRA.07,Sudjana.et.al.PRA.07}. We observe that when Eve uses the minimum required amount of entanglement resources, i.e., $\mathcal{E}(\gamma_{\text{min}})$, the attack reduces to the optimal individual attack. At this point, the beam-splitter $B_{\eta}$ is not interacting with the signal, i.e., $\eta=1$, and the parameter $\kappa$ becomes irrelevant.

Eve's information monotonically increases with the amount of entanglement, and it approaches the Holevo bound in the limit of infinite entanglement, i.e., $\mathcal{E}(\gamma {\rightarrow} 1) \rightarrow \infty$. In this extreme point, the beam-splitter $B_{\eta}$ has transmissivity equal to the channel's transitivity, i.e., $\eta=\tau$, and the state $\boldsymbol{\varphi}$ has a squeezing parameter $\kappa=\sqrt{(\varepsilon-1)/(\varepsilon+1)}$. It is worth noting that the teleportation part of the protocol at this stage is operating under the Choi-state \cite{Holevo.JMP.11} (maximally entangled state sent through the channel). So, having access to a maximally entangled state is the trade-off in order to reach optimality in the teleportation-based attack without purifying the system.

The notion of optimality in the extreme case of infinite entanglement is justified by the fact that a physical bound is reached i.e., the Holevo bound, that we cannot surpass. A meaningful question to ask at this point may be the following: \textit{for a given finite amount of entanglement what is the optimal collective attack?} To this day, a physical limit that upper bounds the amount of classical information that can be extracted from a quantum channel under the use of finite entanglement has not been established. Thus, even though numerical searches indicate the scheme proposed in this work seems to operate optimally for any value of entanglement, we will forgo making such a strong claim.

Finally, the fact that Eve needs an extremely large amount of entanglement in order to approach the optimal collective attack showcases the robustness of CV-QKD protocols. Interestingly, even the minimum amount of entanglement given in Eq.~(\ref{bound}) is arguably beyond current technological capabilities \cite{Vahlbruch.et.al.PRL.16}. In DV protocols there is no need for a similar analysis, since Bell states are not unphysical, and the entangling cloner can operate without any unrealistic assumptions. 

Another possibility for Eve that it is worth to be investigated would be to use the hybrid type of teleportation introduced in Ref.~\cite{Andersen.Ralph.PRL.13}. In this scheme a continuous variable state splits up to $N$ qubits, that can probabilistically be teleported through DV teleportation \cite{Brassard.Braunstein.Cleve.PD.98}. In this scenario the need for infinite amount of entanglement of a single state is compensated with the need of infinite copies of Bell states. With any finite amount of splitting though this protocol simulates a non-Gaussian channel due to the inevitable truncation of the initial state.

\section{Conclusions}
\label{secc}

In this work, we showed that optimal collective attacks in CV-QKD are always based on an extremely strong assumption that takes different forms depending on the way we model the eavesdropper, i.e., full-system purification, simulation of an identity channel, access to a resource state with infinite amount of entanglement, access to infinite copies of Bell states. However, the requirement of having access to a resource state with infinite amount of entanglement can be ``tamed", and a teleportation-based scheme can be modeled that operates in the regime between the optimal individual and optimal collective attacks depending on the available entanglement resources. Thus, the all-optical teleportation attack we introduced in this paper can be thought of as a universal eavesdropping scheme for Gaussian QKD, that can be reduced to either optimal individual or optimal collective attack depending on the available entanglement resources, without assuming Eve has access to the entire environment. 

An interesting extension of this work would be the analysis of the all-optical teleportation attack in other CV-QKD protocols such as two-way \cite{Pirandola.et.al.NP.08} or discretely-modulated \cite{Namiki.Hirano.PRA.03,Namiki.Hirano.PRL.04} CV-QKD. The main technical hurdle in analyzing the latter case is that although optimal strategies have been found for distinguishing discrete sets of CV states in the purely individual attack regime \cite{Croke.Barnett.AOP.09}, finding the optimal collective attack for such non-Gaussian protocols is much more challenging although recently progress has been made in this area \cite{Lin.Upadhyaya.Lutkenhaus.arxiv.19,Kaur.Guha.Wilde.arxiv.19,Ghorai.et.al.PRX.19}.

\section*{Acknowledgements}

The authors acknowledge useful discussions with Andrew Lance and Thomas Symul. This research was supported by funding from the Australian Department of Defence. This research is also supported by the Australian Research Council (ARC) under the Centre of Excellence for Quantum Computation and Communication Technology (Project No. CE170100012). NW acknowledges funding support from the European Unions Horizon 2020 research and innovation programme under the Marie Sklodowska-Curie grant agreement No.750905 and Q.Link.X from the BMBF in Germany.

\appendix
\renewcommand{\thefigure}{A\arabic{figure}}

\section{Quantum Teleportation}
\label{seca}

Quantum teleportation is one of the key tools in quantum information theory, initially introduced for discrete variables \cite{Bennett.et.al.PRL.93,Brassard.Braunstein.Cleve.PD.98}, and then extended to CV systems \cite{Vaidman.PRA.94,Braunstein.Kimble.PRL.98,Ralph.OL.99} (see also \cite{Braunstein.et.al.PRL.00} for a universal approach on teleportation and Ref.~\cite{Pirandola.et.al.NP.15} for recent advances).There are also protools for hybrid situations \cite{Andersen.Ralph.PRL.13,Marshall.James.JOSAB}. 

Let us assume that we want to teleport a (null mean valued) single-mode Gaussian state \cite{Holevo.PIT.07,Weedbrook.et.al.RVP.12,Adesso.Ragy.OSID.14} with covariance matrix $\boldsymbol{\sigma}_{\text{in}}$ from one place (laboratory 1) to another (laboratory 2). In Gaussian systems, a necessary quantum resource in order to achieve that is a {two-mode entangled state $\boldsymbol{\rho}$, shared between the two laboratories, i.e., a state with a covariance matrix given by
\begin{equation}
\boldsymbol{\rho}=\begin{bmatrix}
a & \cdot & c & \cdot \\
\cdot & a & \cdot & -c \\
c & \cdot & b & \cdot \\
\cdot & -c & \cdot & b 
\end{bmatrix} \,.
\label{resource}
\end{equation}

The quality of the teleportation protocol is limited by the amount of entanglement that is pre-shared between the two parties, and perfect teleportation is achieved only in the limit of maximum amount of entanglement resources. In CV systems, though, maximum entanglement is unphysical, since that would require infinite energy. Thus, realistically, instead of achieving a perfect state teleportation, we always end up with a slightly noisy copy of the target state. This process can be modeled as the decoherence that a quantum channel (completely positive trace-preserving map) induces to a transmitted state \cite{Ralph.Lam.Polkinghorne.JOB.99,Bowen.Bose.PRL.01}. 

In general, any Gaussian channel can be simulated via a quantum teleportation protocol using an appropriate resource state \cite{Giedke.Cirac.PRA.02,Niset.Fiurasek.Cerf.PRL.09,Pirandola.et.al.NC.17}. Given a Gaussian phase-insensitive channel $\mathcal{G}$ the set of resource states that can simulate it have been calculated in Ref.~\cite{Tserkis.Dias.Ralph.PRA.18}, generalizing previous results \cite{Scorpo.et.al.PRL.17}.

\subsubsection{Standard CV Teleportation Protocol}

The most well-known CV teleportation protocol was proposed by Braunstein and Kimble \cite{Braunstein.Kimble.PRL.98}. In this protocol one arm of the resource state is mixed with the input state through a balanced beam-splitter in laboratory 1, followed by a Bell-type measurement, i.e., dual homodyne detection (measuring the $\hat{x}$ quadrature on one arm and the $\hat{p}$ on the other), HD, and the results are sent to laboratory 2 through a classical channel, $\mathcal{CC}$. Finally, in laboratory 2, a displacement operation proportional to the results of these measurements, $\mathcal{D}$, is applied to the other arm of the resource state in order to reconstruct the input state, i.e., teleport it. Graphically this protocol is depicted in Fig.~\ref{figapp}~(a). With a Gaussian resource state of the form of Eq.~(\ref{resource}), this teleportation protocol corresponds to a Gaussian phase-insensitive channel with transmissivity $\tau_{\text{tel}}$ and noise $v_{\text{tel}}$ given by
\begin{subequations}
\begin{gather}
\tau_{\text{tel}} = \lambda \,, \\ 
v_{\text{tel}} = a \lambda - 2 c \sqrt{\lambda}+b \,,
\label{BKchannel}
\end{gather}
\end{subequations}
where $\lambda \geqslant 0$ is the experimentally accessible gain. Note that we assumed an infinite-energy limit in the Bell-type measurement detection.

\begin{figure}[t]
\setcounter{figure}{0}
\centering
  \includegraphics[width=\columnwidth]{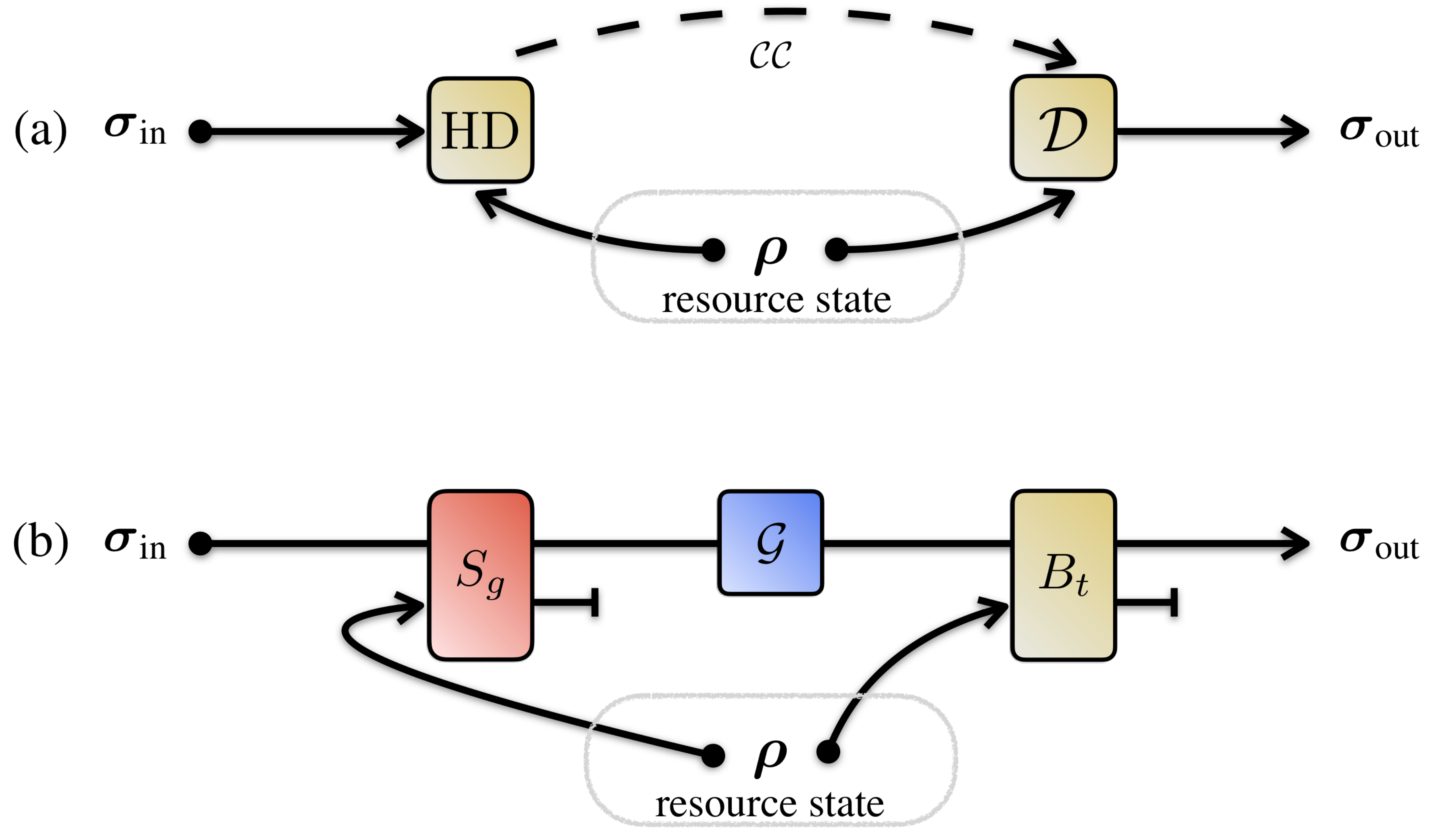}
  \caption{ \small Quantum teleportation protocols. In figure (a) we present the standard CV teleportation protocol via its basic components: (i) the dual homodyne detection, HD, between the resource state, $\boldsymbol{\rho}$, and the initial state $\boldsymbol{\sigma_{\text{in}}}$, (ii) the classical channel, $\mathcal{CC}$, (iii) the displacement, $\mathcal{D}$, and iv) the output state, $\boldsymbol{\sigma_{\text{out}}}$. In figure (b) the all-optical teleportation protocol is presented. In this protocol, the basic components are: (i) a two-mode squeezer $S_g$, (ii) a beam-splitter $B_t$, and (iii) the decoherence that the signal has to go through modeled with a quantum channel $\mathcal{G}$.}
  \label{figapp}
\end{figure}
 
\subsubsection{All-optical Teleportation Protocol}

An alternative all-optical teleportation protocol was proposed by Ralph \cite{Ralph.OL.99}, and it is graphically presented in Fig.~\ref{figapp}~(b). In this protocol one arm of the resource state is fed into a parametric amplifier along with the input state in laboratory 1. The output amplified signal is directly sent to laboratory 2, which means that it has to go through some decoherence that can be modeled as a quantum channel $\mathcal{G}$, that is initially shared between the two laboratories. Finally, in laboratory 2, we mix the signal with the second arm of the resource state on a beam-splitter, where the induced transmissivity is inversely proportional to the amplification applied in laboratory 1. 

Let us assume that we want to teleport the same state $\boldsymbol{\sigma}_{\text{in}}$ as in the standard CV teleportation protocol. In order to do so, we use a two-mode Gaussian resource state with a covariance matrix $\boldsymbol{\rho}$ of the form of Eq.~(\ref{resource}). The initial state can be represented by a combined four-mode covariance matrix $\boldsymbol{\sigma}_{\text{in}} \oplus \boldsymbol{\rho}$. The amplification is achieved by a two-mode squeezer \cite{Caves.PRD.82}, where the two inputs are the initial state and one arm of the resource entangled state. The corresponding symplectic transformation $S_g$ that the amplifier induces is given by
\begin{equation}
S_g=
\left[
\begin{array}{c|c}
\begin{matrix}
\sqrt{g} & \cdot &  \sqrt{g {-}1}  & \cdot  \\
 \cdot & \sqrt{g} & \cdot & - \sqrt{g {-}1}  \\
 \sqrt{g {-}1} & \cdot & \sqrt{g} & \cdot  \\
 \cdot & - \sqrt{g {-}1}  & \cdot & \sqrt{g}
\end{matrix} & \cdot \\
\hline
\cdot & \begin{matrix}
1 & \cdot \\
\cdot & 1
\end{matrix}
\end{array}
\right] \,,
\end{equation}
where $g=\cosh^2 r \geqslant 1$, with $r \in \mathbb{R}$ being the two-mode squeezing parameter. Note that the symplectic transformation $S_g$ is applied to the initial four-mode state $\boldsymbol{\sigma}_{\text{in}} \oplus \boldsymbol{\rho}$, where the identity sub-matrix of $S_g$ indicates that the second arm of the resource state remains unaffected at this stage. 

Applying enough amplification to surpass the quantum limit we end up with a signal that can directly be sent to the other laboratory, but it still needs to go through some decoherence due to the environment. Let us assume that this decoherence is a thermal channel $\mathcal{G}$, with transmissivity/gain $\tau$ and noise $v$.

The subsequent attenuation can be modeled with a beam-splitter, where in one port we feed in the previously amplified state (that is decohered through the environment) and in the other the second arm of the resource entangled state. The symplectic transformation of the beam-splitter $B_t$ is given by
\begin{equation}
B_t=
\left[
\begin{array}{c|c|c}
\begin{matrix}
\sqrt{t } & \cdot  \\
\cdot & \sqrt{t }
\end{matrix} & \cdot & \begin{matrix}
-\sqrt{1{-}t } & \cdot  \\
\cdot & -\sqrt{1{-}t }
\end{matrix}  \\
\hline
\cdot & \begin{matrix}
1 & \cdot \\
\cdot & 1
\end{matrix} & 
\cdot \\
\hline
 \begin{matrix}
\sqrt{1{-}t } & \cdot  \\
\cdot & \sqrt{1{-}t }
\end{matrix} & \cdot & \begin{matrix}
\sqrt{t } & \cdot  \\
\cdot & \sqrt{t }
\end{matrix}
\end{array}
\right] \,,
\end{equation}
with a transmission ratio equal to $t=\lambda/(g \tau)$. Applying on the initial state $\boldsymbol{\sigma}_{\text{in}} \oplus \boldsymbol{\rho}$ the two-mode squeezer $S_g$ we get the amplified state $S_g (\boldsymbol{\sigma}_{\text{in}} \oplus \boldsymbol{\rho}) S_g^T$. This amplified state transforms according to Eq.~(\ref{channel}) due to the decoherence into $\mathcal{G} [S_g (\boldsymbol{\sigma}_{\text{in}} \oplus \boldsymbol{\rho}) S_g^T]$. Finally, this decohered state goes through the final beam-splitter $B_t$ and evolves into $B_t \{ \mathcal{G} [S_g (\boldsymbol{\sigma}_{\text{in}} \oplus \boldsymbol{\rho}) S_g^T] \} B_t^T$. Tracing out mode 2 (the second output of the amplification) and mode 3 (the second output of the attenuation) from this state (see Fig.~\ref{figapp}) we get the output state $\boldsymbol{\sigma}_{\text{out}} = \tr_{23}  B_t \{ \mathcal{G} [S_g (\boldsymbol{\sigma}_{\text{in}} \oplus \boldsymbol{\rho}) S_g^T] \} B_t^T $.

This teleportation protocol corresponds to a Gaussian phase-insensitive channel with transmissivity $\tau_{\text{tel}}$ and noise $v_{\text{tel}}$ given by
\begin{subequations}
\begin{gather}
\tau_{\text{tel}} = \lambda \,, \\
v_{\text{tel}} = a \lambda-2 c \sqrt{\frac{\lambda (g-1) (g \tau  -\lambda )}{ \tau g^2}}-\frac{\lambda  (a \tau +b-v)}{\tau g  }+b  \,.
\label{AOchannel}
\end{gather}
\end{subequations}

In the limit of infinite amplification, i.e., $g\rightarrow \infty$, Eq.~(\ref{AOchannel}) reduces to Eq.~(\ref{BKchannel}) and the output signal of the all-optical teleportation protocol becomes equivalent to that of the standard CV teleportation protocol. For finite amplification and the same amount of entanglement resources, the all-optical teleportation protocol will always correspond to an equally or a slightly more noisy effective channel than the standard CV teleportation protocol, but its big advantage is that there is no need for individual Bell-type measurements during the teleportation process. The significance of this advantage is crucial for the results of the main text.

It is worth noting that in the main text we use the all-optical protocol to simulate the exact same channel as the one that represents the environment, $\mathcal{G}$. Thus, we set $\lambda=\tau$, which implies that $t=1/g$.


\begin{thebibliography}{99}

\bibitem{Scarani.et.al.RMP.09}
V. Scarani et  al., \href{https://journals.aps.org/rmp/abstract/10.1103/RevModPhys.81.1301}{Rev. Mod. Phys. \textbf{81}, 1301 (2009)}.

\bibitem{Pirandola.et.al.arxiv.19}
S. Pirandola et al., \href{https://arxiv.org/abs/1906.01645}{arXiv:1906.01645}.

\bibitem{Xu.et.al.arxiv.19}
F. Xu, X. Zhang, H.-K. Lo, J.-W. Pan, \href{https://arxiv.org/abs/1903.09051}{arXiv:1903.09051}.

\bibitem{Bennett.Brassard.IEEE.84}
C. Bennett and G. Brassard, Proceedings of IEEE, New York, 582 p.175, (1984).

\bibitem{Ekert.PRL.91}
A. K. Ekert, \href{https://journals.aps.org/prl/abstract/10.1103/PhysRevLett.67.661}{Phys. Rev. Lett. \textbf{67}, 661 (1991)}.

\bibitem{Ralph.PRA.99}
T. C. Ralph, \href{https://journals.aps.org/pra/abstract/10.1103/PhysRevA.61.010303}{Phys. Rev. A \textbf{61}, 010303(R) (1999)}.

\bibitem{Hillery.PRA.00}
M. Hillery, \href{https://journals.aps.org/pra/abstract/10.1103/PhysRevA.61.022309}{Phys. Rev. A \textbf{61}, 022309 (2000)}.

\bibitem{Reid.PRA.00}
M. D. Reid, \href{https://journals.aps.org/pra/abstract/10.1103/PhysRevA.62.062308}{Phys. Rev. A \textbf{62}, 062308 (2000)}.

\bibitem{Jouguet.el.al.PRA.11}
P. Jouguet, S. Kunz-Jacques, and A. Leverrier, \href{https://journals.aps.org/pra/abstract/10.1103/PhysRevA.84.062317}{Phys. Rev. A \textbf{84}, 062317 (2011)}.

\bibitem{Jouguet.et.al.NP13}
P. Jouguet, S. Kunz-Jacques, A. Leverrier, P. Grangier, and E. Diamanti, \href{https://www.nature.com/articles/nphoton.2013.63}{Nat. Photonics \textbf{7}, 378 (2013)}. 

\bibitem{Huang.et.al.SR.16}
D. Huang, P. Huang, D. Lin, and G. Zeng, \href{https://www.nature.com/articles/srep19201}{Scientific Reports \textbf{6}, 19201 (2016)}.

\bibitem{Zhang.et.al.QST.19}
Y. Zhang et al., \href{https://iopscience.iop.org/article/10.1088/2058-9565/ab19d1}{Quantum Sci. Technol. \textbf{4}, 035006 (2019)}.

\bibitem{Zhang.et.al.NP.19}
G. Zhang et al., \href{https://www.nature.com/articles/s41566-019-0504-5}{Nat. Photonics \textbf{7}, 378 (2019)}.

\bibitem{Renner.PhD.05}
R. Renner, \href{https://arxiv.org/abs/quant-ph/0512258}{PhD Thesis. ETH Zurich, 2005}.

\bibitem{Holevo.PIT.73}
A. S. Holevo, \href{http://www.mathnet.ru/php/archive.phtml?wshow=paper&jrnid=ppi&paperid=892&option_lang=eng}{Probl. Inf. Transm. \textbf{9}, 31 (1973)}.

\bibitem{Ralph.OL.99}
T. C. Ralph, \href{https://www.osapublishing.org/ol/abstract.cfm?uri=ol-24-5-348}{Opt. Lett. \textbf{24}, 348 (1999)}.

\bibitem{Grosshans.et.al.Nature.03}
F. Grosshans, G. V. Assche, J. Wenger, R. Brouri, N. J. Cerf, and P. Grangier, \href{https://www.nature.com/articles/nature01289} {Nature \textbf{421}, 238-241 (2003)}.

\bibitem{Grosshans.et.al.QIC.03}
F. Grosshans, N. J. Cerf, J. Wenger, R. Tualle-Brouri, and P. Grangier, \href{https://arxiv.org/pdf/quant-ph/0306141.pdf}{Quantum Inf. Comput. \textbf{3}, 535 (2003)}.

\bibitem{Pirandola.Braunstein.Lloyd.PRL.08} S. Pirandola, S. L. Braunstein, and S. Lloyd, \href{https://journals.aps.org/prl/abstract/10.1103/PhysRevLett.101.200504} {Phys. Rev. Lett. \textbf{101}, 200504 (2008)}.

\bibitem{Bennett.et.al.PRL.93}
C.H. Bennett, G. Brassard, C. Crepeau, R. Jozsa, A. Peres, and W. K. Wootters, \href{https://journals.aps.org/prl/abstract/10.1103/PhysRevLett.70.1895}{Phys. Rev. Lett. \textbf{70}, 1895 (1993)}.

\bibitem{Brassard.Braunstein.Cleve.PD.98}
G. Brassard, S. L. Braunstein, and R. Cleve, \href{https://linkinghub.elsevier.com/retrieve/pii/S0167278998000438}{Physica D, \textbf{120}, 43-47 (1998)}.

\bibitem{Vaidman.PRA.94}
L. Vaidman, \href{https://journals.aps.org/pra/abstract/10.1103/PhysRevA.49.1473}{Phys. Rev. A \textbf{49}, 1473 (1994)}.

\bibitem{Braunstein.Kimble.PRL.98}
S. L. Braunstein and H. J. Kimble, \href{https://journals.aps.org/prl/abstract/10.1103/PhysRevLett.80.869}{Phys. Rev. Lett. \textbf{80}, 869 (1998)}.

\bibitem{Braunstein.et.al.PRL.00}
S. L. Braunstein, G. M. D'Ariano, G. J. Milburn, and M. F. Sacchi, \href{https://journals.aps.org/prl/abstract/10.1103/PhysRevLett.84.3486}{Phys. Rev. Lett. 84, 3486 (2000)}.

\bibitem{Andersen.Ralph.PRL.13}
U. L. Andersen and T. C. Ralph, \href{https://journals.aps.org/prl/abstract/10.1103/PhysRevLett.111.050504}{Phys. Rev. Lett. \textbf{111}, 050504 (2013)}.

\bibitem{Marshall.James.JOSAB.14}
K. Marshall and D. F. V. James, \href{https://www.osapublishing.org/josab/abstract.cfm?uri=josab-31-3-423}{J. Opt. Soc. Am. B \textbf{31}, 423 (2014)}.

\bibitem{Pirandola.et.al.NP.15}
S. Pirandola, J. Eisert, C. Weedbrook, A. Furusawa, and S. L. Braunstein \href{https://www.nature.com/articles/nphoton.2015.154}{Nat. Photonics \textbf{9}, 641-652 (2015)}.

\bibitem{Lodewyck.Grangier.PRA.07}
J. Lodewyck and P. Grangier, \href{https://journals.aps.org/pra/abstract/10.1103/PhysRevA.76.022332}{Phys. Rev. A \textbf{76}, 022332 (2007)}.

\bibitem{Sudjana.et.al.PRA.07}
J. Sudjana, L. Magnin, R. Garcia-Patron, and N. J. Cerf, \href{https://journals.aps.org/pra/abstract/10.1103/PhysRevA.76.052301}{Phys. Rev. A \textbf{76}, 052301 (2007)}.

\bibitem{Hosseinidehaj.Walk.Ralph.PRA.19}
N. Hosseinidehaj, N. Walk, and T. C. Ralph, \href{https://journals.aps.org/pra/abstract/10.1103/PhysRevA.99.052336}{Phys. Rev. A \textbf{99}, 052336 (2019)}.

\bibitem{Pan.et.al.arXiv.19}
Z. Pan, K. P. Seshadreesan, W. Clark, M. R. Adcock, I. B. Djordjevic, J. H. Shapiro, and S. Guha, \href{https://arxiv.org/abs/1903.03136}{arXiv:1903.03136}.

\bibitem{Garcia-Patron.PhD.07}
R. Garc\'ia-Patr\'on, \href{http://quic.ulb.ac.be/_media/publications/2007-thesis-raul.pdf}{PhD Thesis. Universite Libre de Bruxelles, 2007}.

\bibitem{Weedbrook.et.al.RVP.12}
C. Weedbrook, S. Pirandola, R. Garc\'ia-Patr\'on, N. J. Cerf, T. C. Ralph, J. H. Shapiro, and S. Lloyd, \href{https://journals.aps.org/rmp/abstract/10.1103/RevModPhys.84.621}{Rev. Mod. Phys. \textbf{84}, 621-699 (2012)}.

\bibitem{Cerf.Levy.VanAssche.PRA.01} N. J. Cerf, M. Levy, and G. Van Assche, \href{https://journals.aps.org/pra/abstract/10.1103/PhysRevA.63.052311} {Phys. Rev. A \textbf{63}, 052311 (2001)}.

\bibitem{Grosshans.Grangier.PRL.02} F. Grosshans and P. Grangier, \href{https://journals.aps.org/prl/abstract/10.1103/PhysRevLett.88.057902} {Phys. Rev. Lett. \textbf{88}, 057902 (2002)}.

\bibitem{Renner.Cirac.PRL.2009} R. Renner and J. I. Cirac, \href{https://journals.aps.org/prl/abstract/10.1103/PhysRevLett.102.110504}{Phys. Rev. Lett. \textbf{102}, 110504 (2009)}.

\bibitem{Garcia-Patron.Cerf.PRL.06}
R. Garc\'ia-Patr\'on, and N. J. Cerf, \href{https://journals.aps.org/prl/abstract/10.1103/PhysRevLett.97.190503}{Phys. Rev. Lett. \textbf{97}, 190503 (2006)}.

\bibitem{Navascues.Grosshans.PRL.06}
M. Navascues, F. Grosshans, and A. Acin, \href{https://journals.aps.org/prl/abstract/10.1103/PhysRevLett.97.190502}{Phys. Rev. Lett. \textbf{97,} 190502 (2006)}.

\bibitem{Leverrier.Grangier.PRA.10}
A. Leverrier, and P. Grangier, \href{https://journals.aps.org/pra/abstract/10.1103/PhysRevA.81.062314}{Phys. Rev. A \textbf{81}, 062314 (2010)}.

\bibitem{Renner.Gisin.Kraus.PRA.05}
R. Renner, N. Gisin, and B. Kraus, \href{https://journals.aps.org/pra/abstract/10.1103/PhysRevA.72.012332}{Phys. Rev. A \textbf{72}, 012332 (2005)}. 

\bibitem{Devetak.Winter.PRSA.05}
I. Devetak, and A. Winter, \href{https://royalsocietypublishing.org/doi/10.1098/rspa.2004.1372}{Proc. R. Soc. A \textbf{461}, 207 (2005)}.

\bibitem{Holevo.B.19}
A. S. Holevo, \textit{Quantum Systems, Channels, Information} (De Gruyter, 2019).

\bibitem{Adesso.Ragy.OSID.14}
G. Adesso, S. Ragy, and A. R. Lee, \href{http://www.worldscientific.com/doi/abs/10.1142/S1230161214400010}{Open Syst. Inf. Dyn. \textbf{21}, 1440001 (2014)}.

\bibitem{Serafini.B.17}
A. Serafini, \textit{Quantum Continuous Variables: A Primer of Theoretical Methods}, (CRC Press, Boca Raton, FL, 2017).

\bibitem{Duan.et.al.PRL.00}
L.-M. Duan, G. Giedke, J. I. Cirac \& P. Zoller \href{https://journals.aps.org/prl/abstract/10.1103/PhysRevLett.84.2722}{Phys. Rev. Lett. \textbf{84,} 2722 (2000)}.

\bibitem{Simon.PRL.00}
R. Simon \href{https://journals.aps.org/prl/abstract/10.1103/PhysRevLett.84.2726}{Phys. Rev. Lett. \textbf{84,} 2726 (2000)}.

\bibitem{Holevo.PIT.07}
A. S. Holevo, \href{https://link.springer.com/article/10.1134/S0032946007010012}{Probl. Inf. Transm. \textbf{43}, 1 (2007)}.

\bibitem{Namiki.Hirano.PRL.04}
R. Namiki and T. Hirano, \href{https://journals.aps.org/prl/abstract/10.1103/PhysRevLett.92.117901}{Phys. Rev. Lett. \textbf{92,} 117901 (2004)}.

\bibitem{Giedke.Cirac.PRA.02}
G. Giedke and J.I. Cirac, \href{https://journals.aps.org/pra/abstract/10.1103/PhysRevA.66.032316}{Phys. Rev. A \textbf{66}, 032316 (2002)}.

\bibitem{Niset.Fiurasek.Cerf.PRL.09}
J. Niset, J. Fiur\'a\v sek, and N. J. Cerf, \href{https://journals.aps.org/prl/abstract/10.1103/PhysRevLett.102.120501}{Phys. Rev. Lett. \textbf{102}, 120501 (2009)}.

\bibitem{Pirandola.et.al.NC.17}
S. Pirandola, R. Laurenza, C. Ottaviani and L. Banchi, \href{https://www.nature.com/articles/ncomms15043}{Nat. Comm. \textbf{8,} 1 (2017)}.

\bibitem{Tserkis.Dias.Ralph.PRA.18}
S. Tserkis, J. Dias, and T. C. Ralph, \href{https://journals.aps.org/pra/abstract/10.1103/PhysRevA.98.052335}{Phys. Rev. A \textbf{98}, 052335 (2018)}.

\bibitem{Scorpo.et.al.PRL.17}
P. Liuzzo-Scorpo, A. Mari, V. Giovannetti and G. Adesso, \href{https://journals.aps.org/prl/abstract/10.1103/PhysRevLett.119.120503}{Phys. Rev. Lett. \textbf{119}, 120503 (2017)}; P. Liuzzo-Scorpo, A. Mari, V. Giovannetti and G. Adesso, \href{https://journals.aps.org/prl/abstract/10.1103/PhysRevLett.120.029904}{Phys. Rev. Lett. \textbf{120}, 029904 (2018)}.

\bibitem{Caves.PRD.82}
C. M. Caves, \href{https://journals.aps.org/prd/abstract/10.1103/PhysRevD.26.1817}{Phys. Rev. D \textbf{26}, 1817 (1982)}.

\bibitem{Holevo.Sohma.Hirota.PRA.99}
A. S. Holevo, M. Sohma, and O. Hirota \href{https://journals.aps.org/pra/abstract/10.1103/PhysRevA.59.1820}{Phys. Rev. A \textbf{59}, 1820-1828 (1999)}.

\bibitem{Bennett.DiVincenzo.et.al.PRA.96}
C. H. Bennett, D. P. DiVincenzo, J. A. Smolin \& W. K. Wooters \href{https://journals.aps.org/pra/abstract/10.1103/PhysRevA.54.3824}{Phys. Rev. A \textbf{54,} 3824 (1996)}.

\bibitem{Leverrier.PRL.15}
A. Leverrier, \href{https://journals.aps.org/prl/abstract/10.1103/PhysRevLett.114.070501}{Phys. Rev. Lett. \textbf{114}, 070501 (2015)}.

\bibitem{Leverrier.PRL.17}
A. Leverrier, \href{https://journals.aps.org/prl/abstract/10.1103/PhysRevLett.118.200501}{Phys. Rev. Lett. \textbf{118}, 200501 (2017)}.

\bibitem{Holevo.JMP.11}
A. S. Holevo, \href{https://aip.scitation.org/doi/10.1063/1.3581879}{J. Math. Phys. \textbf{52}, 042202 (2011)}.

\bibitem{Vahlbruch.et.al.PRL.16}
H. Vahlbruch, M. Mehmet, K. Danzmann, and R. Schnabel, \href{https://journals.aps.org/prl/abstract/10.1103/PhysRevLett.117.110801}{Phys. Rev. Lett. \textbf{117}, 110801 (2016)}.

\bibitem{Marshall.James.JOSAB}
K. Marshall and D. F. V. James, \href{https://www.osapublishing.org/josab/abstract.cfm?uri=josab-31-3-423}{J. Opt. Soc. Am. B \textbf{31}, 423 (2014)}.

\bibitem{Ralph.Lam.Polkinghorne.JOB.99}
T. C. Ralph, P. K. Lam and R. E. S. Polkinghorne, \href{http://iopscience.iop.org/article/10.1088/1464-4266/1/4/321}{J. Opt. B: Quantum Semiclass. Opt. \textbf{1}, 483–489 (1999)}.

\bibitem{Bowen.Bose.PRL.01}
G. Bowen and S. Bose, \href{https://journals.aps.org/prl/abstract/10.1103/PhysRevLett.87.267901}{Phys. Rev. Lett. \textbf{87}, 267901 (2001)}.

\bibitem{Pirandola.et.al.NP.08}
S. Pirandola, S. Mancini, S. Lloyd, and S. L. Braunstein, \href{https://www.nature.com/articles/nphys1018}{Nature Physics, \textbf{4}, 726 (2008)}.

\bibitem{Namiki.Hirano.PRA.03}
R. Namiki and T. Hirano, \href{https://journals.aps.org/pra/abstract/10.1103/PhysRevA.67.022308}{Phys. Rev. A \textbf{67}, 022308 (2003)}.

\bibitem{Namiki.Hirano.PRL.04}
R. Namiki and T. Hirano, \href{https://journals.aps.org/prl/abstract/10.1103/PhysRevLett.92.117901}{Phys. Rev. Lett. 92, 117901 (2004)}.

\bibitem{Croke.Barnett.AOP.09}
S. Croke and S. M. Barnett, \href{https://www.osapublishing.org/aop/abstract.cfm?uri=aop-1-2-238}{Advances in Optics and Photonics, \textbf{1}, 238 (2009)}.

\bibitem{Lin.Upadhyaya.Lutkenhaus.arxiv.19}
J. Lin, T. Upadhyaya, and N. L\"{u}tkenhaus, \href{https://arxiv.org/abs/1905.10896}{arXiv:1905.10896}.

\bibitem{Kaur.Guha.Wilde.arxiv.19}
E. Kaur, S. Guha, M. M. Wilde, \href{https://arxiv.org/abs/1901.10099}{arXiv:1901.10099}.

\bibitem{Ghorai.et.al.PRX.19}
S. Ghorai, P. Grangier, E. Diamanti, and A. Leverrier, \href{https://journals.aps.org/prx/abstract/10.1103/PhysRevX.9.021059}{Phys. Rev. X 9, 021059 (2019)}.

\end{thebibliography}
\end{document}